\documentclass[iop]{emulateapj}
\usepackage{apjfonts}

\usepackage[breaklinks,colorlinks,citecolor=blue]{hyperref}
\usepackage{xcolor}
\usepackage{xspace}

\usepackage{mathtools}
\usepackage{amsmath}
\usepackage{booktabs}

\usepackage{natbib}
\usepackage{tabularx}
\usepackage{threeparttable}

\nocite{}

\def\kpch{{\rm\thinspace kpc/h}\xspace}

\def\Mpch{{\rm\thinspace Mpc/h}\xspace}

\def\Msun{\hbox{$\thinspace M_{\odot}$}\xspace}

\def\Gyr{{\rm\thinspace Gyr}\xspace}
\def\Myr{{\rm\thinspace Myr}\xspace}

\def\sub{_\textrm}

\begin{document}

\title{YZiCS: Preprocessing of dark halos in the hydrodynamic zoom-in simulation of clusters}
\shorttitle{Preprocessing of dark halos}

\author{San Han\altaffilmark{1}, Rory Smith\altaffilmark{2}, Hoseung Choi\altaffilmark{1}, Luca Cortese\altaffilmark{3, 4}, Barbara Catinella\altaffilmark{3, 4}, Emanuele Contini\altaffilmark{1}, Sukyoung K. Yi\altaffilmark{1}}
\affil{\altaffilmark{1}Department of Astronomy, Yonsei University, 50 Yonsei-ro, Seodaemun-gu, Seoul 03722, Republic of Korea}
\affil{\altaffilmark{2}Korea Astronomy \& Space Science Institute, Daejeon 305-348, Republic of Korea}
\affil{\altaffilmark{3}International Centre for Radio Astronomy Research, University of Western Australia, 35 Stirling Highway, Crawley, WA 6009, Australia}
\affil{\altaffilmark{4}ARC Centre of Excellence for All Sky Astrophysics in 3 Dimensions (ASTRO 3D)}

\begin{abstract}

To understand the galaxy population in clusters today, we should also consider the impact of previous environments prior to cluster infall, namely preprocessing. We use YZiCS, a hydrodynamic high-resolution zoom-in simulation of 15 clusters, and focus on the tidal stripping suffered by the dark matter halos of cluster members due to preprocessing. We find $\sim48\%$ of today's cluster members were once satellites of other hosts. This is slightly higher than previous estimates, in part, because we consider not just group-mass hosts, but hosts of all masses also. Thus, we find the preprocessed fraction is poorly correlated with cluster mass and is instead related to each cluster's recent mass growth rate. Hosts less massive than groups are significant contributors, providing more than one-third of the total preprocessed fraction. We find that halo mass loss is a clear function of the time spent in hosts. However, two factors can increase the mass loss rate considerably; the mass ratio of a satellite to its host, and the cosmological epoch when the satellite was hosted. The latter means we may have previously underestimated the role of high redshift groups. From a sample of heavily tidally stripped members in clusters today, nearly three quarters were previously in a host. Thus, visibly disturbed cluster members are more likely to have experienced preprocessing. Being hosted before cluster infall enables cluster members to experience tidal stripping for extended durations compared to direct cluster infall and at earlier epochs when hosts were more destructive.

\end{abstract}

\setcounter{footnote}{4}

\section{Introduction} \label{sec:intro}
It is well known that galaxy morphology has a clear dependence on the environment \citep{Dressler1980, Dressler1997, Miller2003, Goto2003, Balogh2004, Kauffmann2004}. In dense regions (e.g. clusters), galaxies are more likely to be bulge-dominated, have red colors and low star formation rates. In sparse regions (the field and voids), galaxies are mostly disk-dominated, have blue colors and are actively star-forming. This morphology-density relation is commonly attributed to environmental mechanisms which transform star-forming galaxies into passive galaxies inside of clusters.

Various processes that may act on cluster members have been suggested, including: tidal stripping by the gravitational potential of the host \citep{Richstone1976, Gnedin2003, Wetzel2010}, ram-pressure stripping by the hot intracluster medium (ICM) \citep{Gunn1972, Quilis2000, Jachym2007, Tonnesen2007, Boselli2008, McCarthy2008}, and multiple close tidal encounters with other members, known as harassments \citep{Moore1996, Moore1998, Gnedin2003a, Mastropietro2005, Aguerri2009, Smith2015}. However, it is important to note that these terms are not necessarily cluster-exclusive. Some may also be applicable in the environment of groups or even in lower-mass hosts.

While physical mechanisms acting in clusters might explain the high fraction of early-type galaxies found within one virial radius of the cluster center, the fraction of red galaxies continues to increase with density at many virial radii from the cluster, where cluster tides and the ICM density is thought to be too low to directly affect these galaxies \citep{Dressler1980, Cybulski2014}. Some galaxies found beyond the virial radius may have previously been fallen inside the cluster and have since bounced back to its outskirts \citep{Balogh2000, Mamon2004}. These are called `backsplash' galaxies, and they may be difficult to distinguish from a newly infalling population of galaxies, which makes the cluster's influence seem to extend beyond the virial radius \citep{Wetzel2014}. Using N-body simulations, \citet{Gill2005} found that approximately half of the galaxies within the distance range $1\text{--}2$ viral radii are backsplash galaxies. Nevertheless, back-splash galaxies are not expected to extend beyond $\sim$3 viral radii at maximum. And, according to recent studies, the backsplash population cannot fully explain the decreasing slope of star formation rate toward the cluster \citep{Jaffe2015, Haines2015}. One solution to this problem could be the role of non-cluster environmental processes. Less dense structures such as filaments, and in particular groups, may begin transforming galaxies even before they reach the cluster environment. This process is known as galaxy `preprocessing' \citep{Mihos2004, Fujita2004}.

Many observational studies provide supporting evidence for the role of preprocessing \citep{Cortese2006, Nantais2016, Roberts2017, Bianconi2018}. Although mergers are expected to be highly rare inside of clusters due to the high relative velocity ($\sim2000$ km/s) between members \citep{Binney1987, Gnedin2003a}, \citet{Sheen2012} found $\sim25\%$ of bright red sequence galaxies in clusters have features known as `merger relics' (e.g. asymmetric features, shells, rings, dust lanes) which are likely to be post-merger signatures.

Also, \citet{Toloba2014} found several examples of cluster dwarf elliptical galaxies with young counter-rotating cores. Possible processes responsible for the formation of such internal dynamics include mergers \citep{Hernquist1991} and gas accretion \citep{Balcells1990}. Mergers in clusters are expected to be rare as mentioned above. Likewise, direct external gas accretion is nearly impossible inside of clusters due to the presence of a prohibiting hot ICM. Taken together, these features could be considered evidence for preprocessing in the less dense environments prior to infall into the cluster \citep{Yi2013}.

Substructures within a cluster \citep{Knebe2000, Zhang2007, Jaffe2012, Jaffe2013}, consisting of a local concentration of galaxies, and perhaps their own hot gas halo, are considered to be the result of group-cluster mergers. \citet{Mahajan2013} found that post-starburst galaxies in clusters were often members of substructures, and so could be an outcome of preprocessing.

Numerical simulations are a useful tool to study the effects and significance of preprocessing. By matching SDSS galaxies with N-body cosmological simulations using halo abundance matching, \citet{Wetzel2013} demonstrated that group preprocessing must play a critical role in the quenching of galaxies. \citet{Villalobos2012, Villalobos2014} ran an idealized N-body simulation of the group-satellite system to reproduce the tidal stripping and harassment. They found that the mass ratio of a satellite to its group is an important factor influencing the amount of mass loss the satellite suffers, with more massive satellites resulting in a greater mass loss. More recently, using dark matter only cosmological simulations, \citet{Joshi2017} investigated tidal stripping of cluster members. They found that, on average, group members lose $35\text{--}45\%$ of their peak mass prior to cluster infall, suggesting a strong influence of preprocessing for group members. Also, they found that members which fall into the cluster as a single halo `catch-up' with the mass loss of group halos by rapidly losing mass. \citet{Lisker2013} studied the observed morphology-distance relation of dwarf ellipticals in clusters using a semi-analytic model approach and argued that these galaxies are likely to have been affected by the group or cluster environment at earlier epochs. \citet{Vijayaraghavan2013} found that groups contribute to preprocessing by enhancing the merger rate compared to in clusters and that, after cluster infall, the group members stay coherent for a long time, which would be observed as substructures. \citet{Rudick2009} and \citet{Contini2014} have revealed that a large fraction of the intracluster light is composed of stars stripped from violently interacting group galaxies at earlier epochs of the Universe.

Galaxies that were affected by preprocessing may now be mixed in with other members of the cluster today, which makes them difficult to identify. Therefore, quantifying the fraction of cluster members that were formerly members of a group (i.e. the pre-processed fraction) can be a good starting point to estimate the impact of group pre-processing on the cluster population. Using a semi-analytical model approach, \citet{McGee2009} and \citet{DeLucia2012} estimated $\sim28\%$ of the galaxies with a stellar mass larger than $10^{9}$\Msun found today in clusters of mass $10^{14}$\Msun were former group members. The preprocessed fraction was found to vary from $15\%$ to $50\%$ depending on the mass of the clusters, hosts, and satellites \citep{Berrier2009, McGee2009}. These studies considered only hosts more massive than $10^{13}$\Msun as possible sources of pre-processing. However, massive groups may not be the sole driver of preprocessing. There is abundant proof that interactions with hosts that inhabit lower mass halos, such as the Milky Way or M31 system, can cause significant transformation of their satellites \citep{Ibata1994, Ibata2001, Grebel2001, Mayer2006, Nidever2008}. Although it is yet unclear how significant their contribution might be to the cluster population, we have decided not to neglect any mass of host in this study.

Also, even at fixed mass, groups formed at earlier epochs should have distinctive merger trees compared to today's groups, and hence do not share the same progenitors. Today's groups are spawned in much lower environmental densities and take a longer time to assemble. This could potentially cause the effects of preprocessing to depend on redshift. However, observations of high-redshift groups, which form a core part of today's cluster, is highly challenging. Typically, direct witnessing of group preprocessing \citep{Cortese2006, Bianconi2018} is only possible in the nearby universe. As a result, this could lead us to underestimate or misunderstand the role of the group environment in shaping the current day cluster galaxy population.

This analogy is similar to the concept of `\textit{progenitor bias}', which was originally proposed to explain the time evolution of cluster population \citep{vanDokkum1996}. Because star-forming galaxies are continuously transformed, the new members add to the original quenched galaxy population \citep{Lilly2016}. Therefore, in this case, early-type galaxy populations in clusters at high redshift do not have a one-to-one correspondence with today's cluster populations. The term also refers to issues that arise when comparing two galaxy populations that have a similar current day property but arrived at this point along a very different evolutionary path. An example of this is that early-type cluster members with a similar stellar mass to late-type cluster members may have undergone very different star formation histories in order to arrive at their current day mass before their star formation was halted \citep{Mistani2016}.

In this study, we will quantify the significance of preprocessing by measuring the fraction of cluster members today that have suffered preprocessing. We focus on the mass evolution of dark halos and try to determine which parameters control the amount of halo mass loss from preprocessing. We apply a new criterion to identify satellites and consider all masses of halos as possible hosts that can cause preprocessing.

This paper is organized as follows. In Section \ref{sec:methods}, we briefly introduce the simulations and describe the data analysis and satellite selection criterion. In Section \ref{sec:results}, we present the main results and discuss them in Section \ref{sec:discussion}. We conclude in Section \ref{sec:conclusion}.

\section{Methods}\label{sec:methods}
\subsection{Numerical Simulation} \label{sec:simulations}
For this study, we use a hydrodynamic zoom-in simulation of clusters, namely, YZiCS (Yonsei Zoom-in Cluster Simulation) \citep{Choi2017}. These simulations were conducted with the adaptive mesh refinement code, RAMSES \citep{Teyssier2002}. First, a dark matter only cosmological volume simulation, based on WMAP7 cosmological parameters ($h_0=0.704$, $\Omega_m=0.272$, $\Omega_\Lambda=0.728$, $\sigma_8=0.809$, and $\rm n = 0.963$) \citep{Komatsu2011} was carried out in a 200\Mpch box. Several dense environments were then selected and resimulated with a full hydrodynamic zoom-in simulation. The effective spatial resolution of the zoom-in simulation is 0.76\kpch. The mass of each dark matter particle is $7\times10^7\Msun$. Snapshots were obtained at a scale factor interval of 0.005 ($\sim$75\Myr). Then AdaptaHOP \citep{Aubert2004} was used to find halos and galaxies in the zoom-in region and measure their positions, velocities, and masses. Finally, Consistent Trees \citep{Behroozi2013} were used to generate merger trees for both galaxies and dark matter halos.

Our zoom-in simulations have comparable spatial force resolution with recent cosmological hydrodynamic volume simulations \citep{Dubois2014, Vogelsberger2014, Schaye2015}. One advantage is, by conducting zoom-in simulations on a 200\Mpch box, we find significantly more high-density environments and massive clusters that are found in most volume simulations, which are typically limited to a 100\Mpch box size. 

At $z=0$, each zoom-in region contains one main cluster halo, with a mass in the range $10^{13.7}\text{--}10^{15.0}\Msun$. We define these halos as `cluster' halos, and any other halo that contains at least one of its own satellites as a `host halo'. We should be cautious because our choice of which clusters to zoom-on has resulted in a cluster sample that does not tightly follow the universal cluster mass distribution. However, in Section \ref{sec:prep_frac}, we will show that the results of this study are not strongly dependent on cluster mass, so we do not expect there to be a significant bias due to this issue.

Our main focus in this study is to measure the mass evolution of dark matter halos that undergo preprocessing. However, our simulations are fully hydrodynamical, containing both gas, star particles, and complex feedback recipes. In fact, a hydrodynamic simulation is a good choice for studying tidal stripping of dark matter, as a dense stellar component at the center of a host halo will generate enhanced dynamical friction and additional tidal forces that could alter the mass evolution of satellites \citep{Arraki2014}. In addition, the presence of baryons can cause halo contraction, while supernova and AGN feedback have been also demonstrated to modify not only the halo density profiles \citep{Governato2012, Peirani2017}, but also the number of subhalos \citep{Wetzel2016} in the dark matter halos. In addition, we have already studied in detail the manner in which tidal stripping of stars is linked with tidal mass loss of the dark matter halo \citep{Smith2016} using these same simulations. In this previous study, we found that stellar stripping only becomes significant once large quantities ($\sim$80$\%$) of the dark matter halo have been tidally stripped.

Although ours simulation do include stars, we will primarily focus on the dark matter evolution in this study. In this way, our results are not strongly dependent on the exact choice of baryonic physics prescription, and are less affected by the limited spatial resolution of cells surrounding the stellar component.

\subsection{Data Selection}
The detection limit of the dark matter halo in the simulation is 64 particles, which corresponds to a minimum detectable halo mass of $\sim5\times10^{9}\Msun$. However, we take a conservative mass cut at $M\sub{cut} = 10^{11}\Msun$, which is equivalent to $\sim1400$ dark matter particles. This ensures that the halo mass function is little affected by the resolution limit and so we have a complete sample of halos over the entire mass range we consider. It also greatly reduces the number of broken merger trees that arise when halos cross the dense cluster core and could otherwise impact on the statistics of cluster members that are the main focus of this study. We also rejected any trees that appear to start inside the cluster as these are likely to be broken trees. We only consider halos that are cluster members at $z=0$. Our final sample consists of 2387 halos.

\begin{figure}
\figurenum{1}
\phantomsection
\label{fig:f1}
\includegraphics[width=0.5\textwidth]{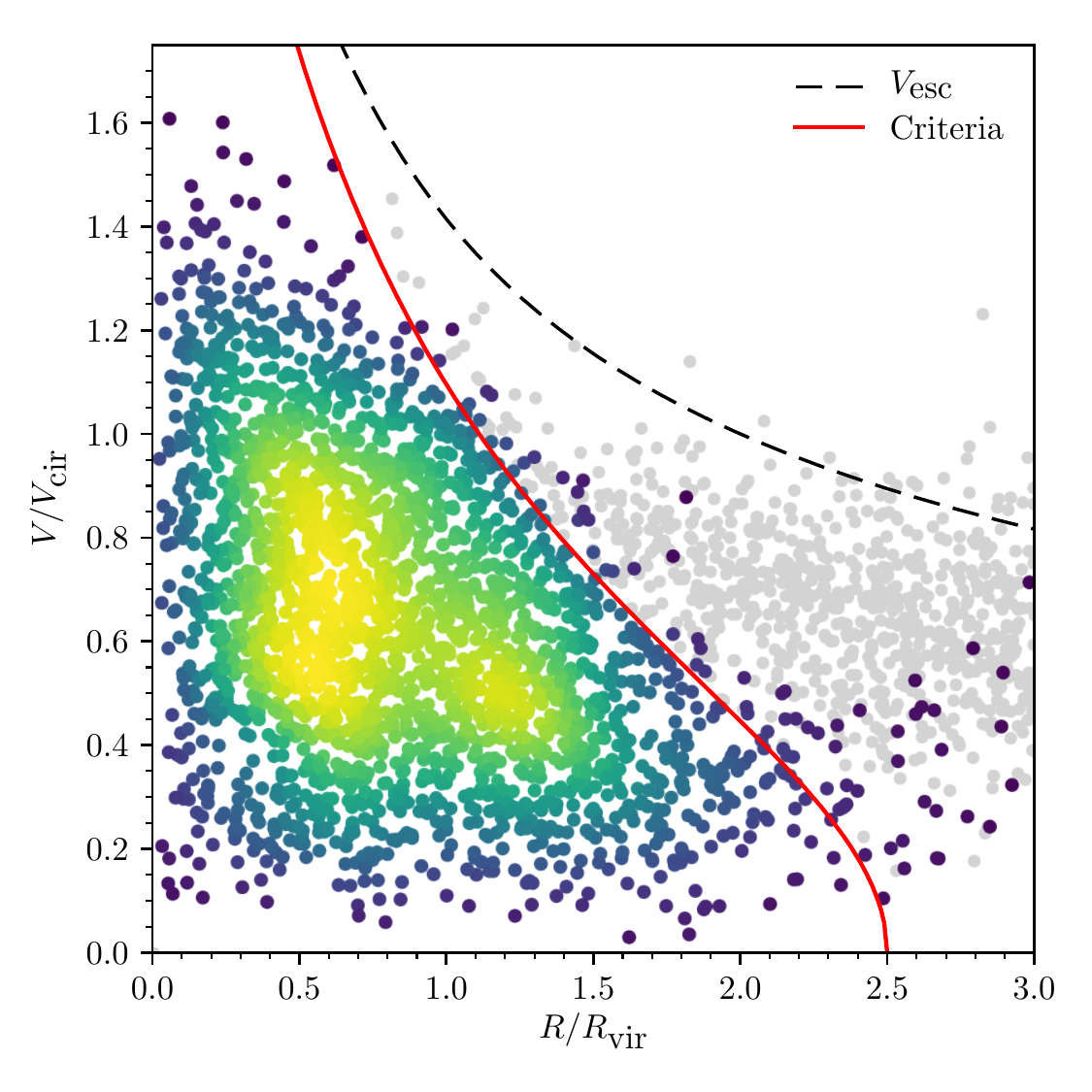}\caption{The solid red line indicates the criterion used to classify satellite members. Points that enter the region beneath the red solid line are tagged as satellites. The y-axis is the 3D orbital velocity with respect to the host normalized by the circular velocity of the host. The x-axis is the 3D radius from the host center normalized by the host virial radius. In this figure, we stack all of the halos from our 15 main clusters at $z=0$ together. Halos that have never met the satellite criterion of their respective main cluster are colored as gray. Halos that have satisfied the satellite criterion at any instant in the past are shown as colored points, where the color indicates their number density.}
\end{figure}

\begin{figure*}[t]
\figurenum{2}
\phantomsection
\label{fig:f2}
\includegraphics[width=1\textwidth]{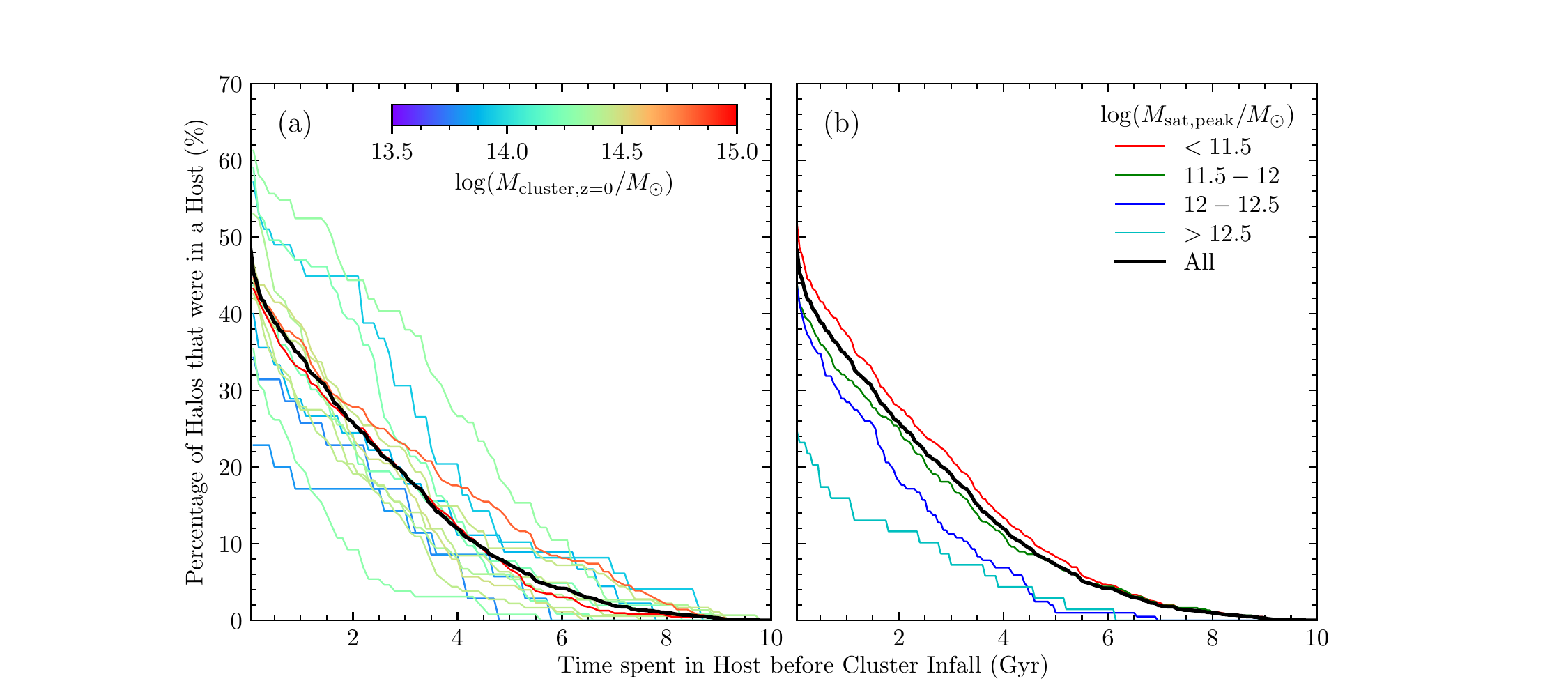}\caption{Cumulative percentage of the cluster members at $z=0$ that have spent more than a certain amount of time in a host before infall. The thick dark line in each panel shows the percentage of all members in all 15 clusters combined. In panel (a), each individual cluster is drawn as a thin line with varying line style (see the legend, for style, cluster id, and $z=0$ cluster mass in log units). In panel (b), the sample is divided by the peak halo mass of the cluster member, and the percentage of members is measured in each mass bin.}
\end{figure*}

\subsection{Satellite membership criterion}
Several issues must be considered when defining satellites of hosts in the simulations. Many studies use the virial radius of the main halo as the outermost boundary of the system \citep{Berrier2009, McGee2009}. However, we did not adopt this approach for following reasons. Firstly, the criterion can mistakenly recognize `fly-bys' as satellites. These are halos that may briefly enter the virial radius of a host, but are never true satellites of that host, as they are traveling at velocities beyond the host's escape velocity. Secondly, it excludes the backsplash population -- these halos pass through their host and exit the host virial radius, often reaching up to $2.5R\sub{vir}$ from the host \citep{Mamon2004} before falling back in. Therefore, by using the virial radius, satellites may appear to repeatedly leave their host before returning sometime later, which was a situation we wished to avoid in this study. Also, satellites may begin to experience environmental effects from their host at distances beyond the host's virial radius, such as tidal stripping \citep{Hahn2009, Rhee2017} or ram pressure stripping \citep{Bahe2013}. In our models, we found that the radius at which we see tidal stripping begins is typically $\sim2R\sub{vir}$, for isolated halos falling into their host for the first time, and we find the result is quite independent of the host mass or mass ratio. Therefore, we devised our own criterion for defining satellite membership, that can extend to as large as $2.5R\sub{vir}$ from the host, but also takes into account the relative velocity of the satellite with respect to its host as an additional parameter. The definition is as follows:

\begin{figure}[t]
\figurenum{3}
\phantomsection
\label{fig:f3}
\includegraphics[width=0.5\textwidth]{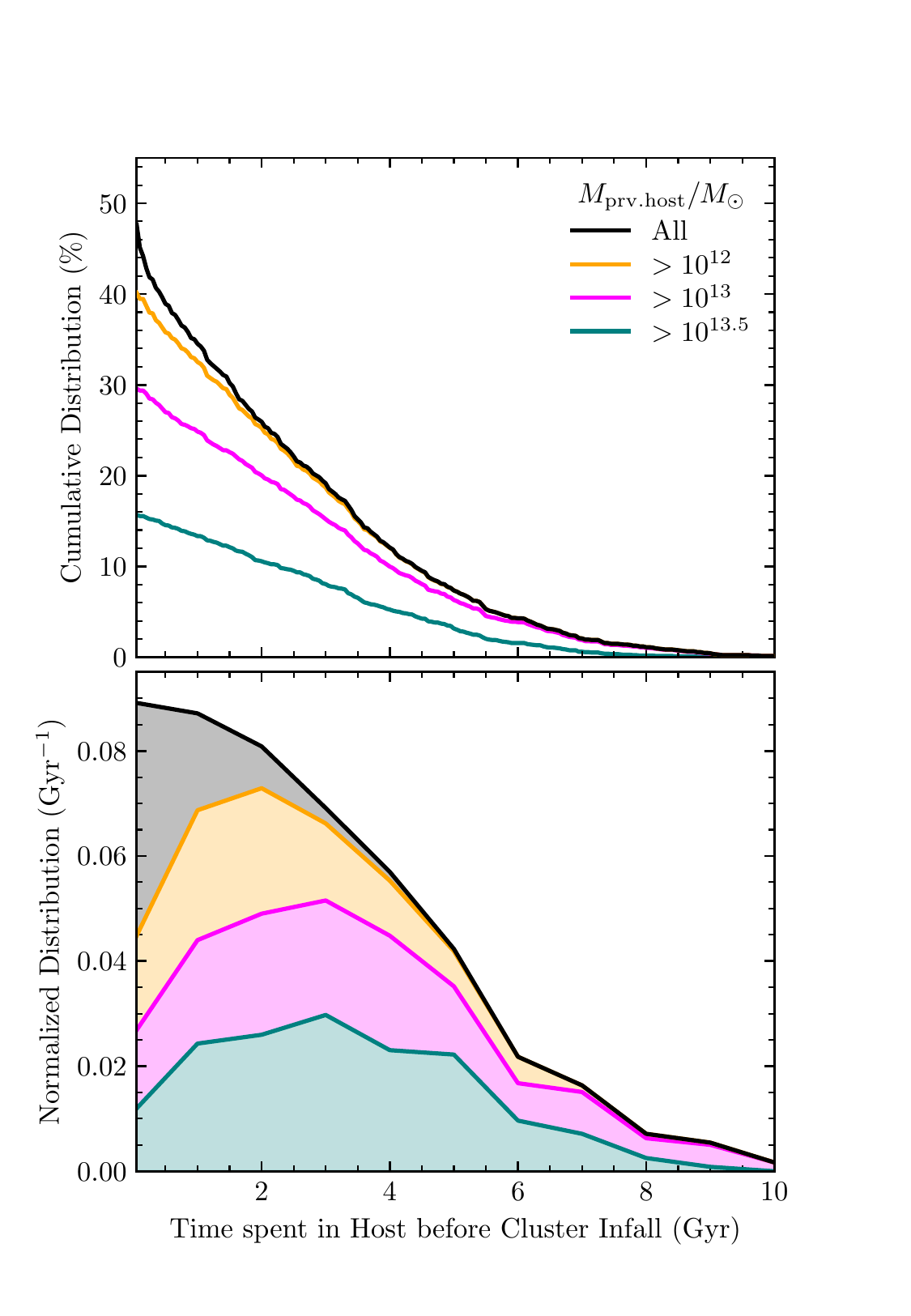}\caption{The relative contribution of hosts with different masses to the preprocessed population. We divide the sample by mass of the host prior to infall into the cluster. In the upper panel, we show the cumulative percentage of cluster members that spent longer than a specified time in hosts. In the lower panel, we show the normalized frequency distribution along each time bin. The mass of the host is measured at its peak. Colored areas (gray, orange, magenta, teal) indicate the fraction of halos that came from each host mass bin, $<10^{12}$, $10^{12}\text{--}10^{13}$, $10^{13}\text{--}10^{13.5}$, and $>10^{13.5}$ in the unit of $\Msun$. Most of the preprocessed population comes from group-mass hosts (the area below the orange line) and the contribution increases to unity as we approach longer times spent in host. Meanwhile, among halos that spent little time in a host, hosts that are less massive than $10^{13}\Msun$ contribute a significant fraction of the preprocessed members.}
\end{figure}

\begin{figure*}
\figurenum{4}
\phantomsection
\label{fig:f4}
\includegraphics[width=1\textwidth]{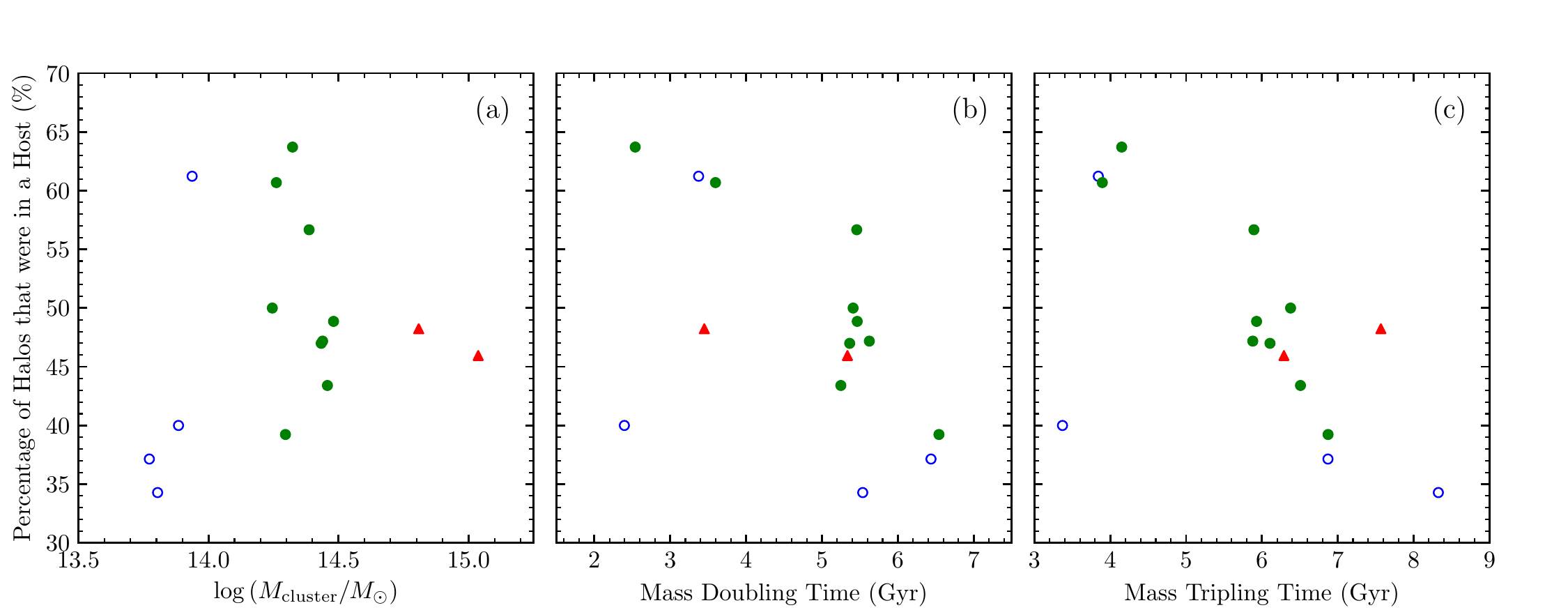}\caption{Percentage of the $z=0$ cluster members that were in a host for at least one snapshot, sub-sampled according to (a) the final virial mass of the cluster, (b) the time required to double their mass and (c) the time required to triple their mass. Symbol color and shape is chosen by the cluster mass, as demonstrated in (a). In (b) and (c), clusters with shorter times (more rapid recent growth) tend to have higher percentages, while there is no clear correlation with cluster mass.}
\end{figure*}

\begin{equation}\label{eq1}
\frac{v^2}{2}+\Phi(r)<\Phi(2.5R\sub{vir}).
\end{equation}

where $\Phi(r)$ is the potential of the target host. $v$ is the relative velocity of the satellite halo with respect to its host. The left-hand side of Equation \ref{eq1} is the specific orbital energy of the satellite in its host. By choosing the left-hand side to be less than the right-hand side, we ensure that the satellite membership criterion cannot extend beyond 2.5R$\sub{vir}$. The potential of the central host is calculated assuming that the radial density profile of their dark matter halos follows an NFW profile

\begin{equation}\label{eq2}
\rho(r)=\frac{\rho\sub{crit}\delta\sub{c}}{(r/R\sub{s})(1+r/R\sub{s})^2}
\end{equation}

\noindent
that is truncated at the virial radius.

\begin{equation}\label{eq3}
\rho'(r)=
\begin{dcases}
\rho(r)-\rho(R\sub{vir}), & \text{for } 0<r\leq R\sub{vir}\\
0, & \text{for } r<R\sub{vir}
\end{dcases}
\end{equation}

Therefore, the potential of the profile is given by

\begin{equation}\label{eq4}
\begin{split}
&\Phi(s)=-V\sub{cir}^2\left[h(c)g(c)\left(\frac{\ln(1+cs)}{s}-\ln(1+c)\right) \right.\\
&\left.\qquad\qquad\qquad\quad-\frac{1}{2}(1-s^2)(h(c)-1)+1\right]\\
\end{split}
\end{equation}

\noindent
where each term is defined by

\begin{equation}\label{eq5}
g(c)\equiv\left[\ln(1+c)-\frac{c}{1+c}\right]^{-1}
\end{equation}

\begin{equation}\label{eq6}
h(c)\equiv\left[1-\frac{c^2g(c)}{3(1+c)^2}\right]^{-1}
\end{equation}

\noindent
and $s=r/R\sub{vir}$ is the distance of satellite from its host (in units of the host virial radius), and $V\sub{cir}=\sqrt{GM\sub{vir}/R\sub{vir}}$ is the circular velocity at the virial radius of the host. The concentration index of the NFW halo is defined as $c=R\sub{vir}/R\sub{s}$.

The resulting criterion is shown by the red line in Figure \ref{fig:f1}. Here, each point represents a single halo, and the halos surrounding the main host in all 15 zoom-in simulations are stacked together. The x-axis shows the distance of each halo from the main host in units of the main host's virial radius. The y-axis is the relative velocity of satellites with respect to their main host in units of the host's circular velocity. Gray points are halos that never have met the host's satellite criterion and tend to lie in the region, indicating they are infalling towards the host for the first time. Colored points are halos that have, at one previous instant, entered beneath the red solid line. As can be seen, the vast majority of colored points are found beneath the solid red line. This means that, once a galaxy has been identified as a member, they tend to remain as a member and continue to obey the criterion. Thus, our criterion effectively selects objects that are associated with their hosts, even if they are currently found beyond the virial radius of their host (i.e. backsplash galaxies). We define the `infall time' as the moment when a halo first becomes a member of its host, following the above criterion for membership. We also note that the same criterion is applied for when halos become members of the cluster, too.

\section{Results}\label{sec:results}
In order to follow the evolution of the individual cluster members, we track the branch of most massive massive progenitor in their merger tree. We measure the total time that a cluster member spent inside another host before becoming a cluster member.

\subsection{Preprocessed fraction}\label{sec:prep_frac}

We find that, on average, $\sim48\%$ of the halos that are cluster members at $z=0$ were previously in a host in at least one snapshot ($\sim75\Myr$). We define this population of halos as the `preprocessed fraction'. Figure \ref{fig:f2} shows the cumulative percentage of halos that have stayed in a host for more than a specified amount of time before becoming a cluster member. In panel (a), we look at cluster-to-cluster variations. Each thin colored line represents an individual cluster and the color shows their mass at $z=0$. The thick dark line is the mean value for all the clusters combined. Although almost half of the cluster members have spent time in a host, most of them did not spend much time there. For example, only $\sim12\%$ of the cluster members have spent more than 4\Gyr in the group. This highlights the fact that, while the cluster accretes new members with time, other hosts are also doing the same prior to infall into the cluster. As a result, many of their members may have recently joined their host and spent only a limited amount of time in the host before joining the cluster population, and this is further enhanced if the group should fall into the cluster at early times. We now consider cluster-to-cluster variations. The percentage cluster members that were former members of hosts varies quite widely (from $\sim23\%$ to $\sim58\%$ depending on the cluster). We will consider the parameters controlling this percentage in the next figure - here we simply note that it is not a clear function of cluster mass.

In panel (b), we present the dependence of the preprocessed fraction on the mass of the cluster member halos (measured at their peak value). The fraction weakly changes with halo mass below $M\sub{peak}\sim10^{12}\Msun$ and drops quickly above $M\sub{peak}\sim10^{12.5}\Msun$. This is because it becomes increasingly difficult to find halos that could act as hosts as they are limited to a narrow mass range -- more massive than the satellite, but less massive than the cluster. The trend seems broadly consistent with the results of \citet{DeLucia2012}, who found a preprocessed fraction of 48\%, 43\%, 23\% for galaxies with stellar masses of $\sim10^{9}\Msun$, $\sim10^{10}\Msun$, $\sim10^{11}\Msun$, corresponding to peak halo masses of approximately $10^{11}\Msun$, $10^{11.5}\Msun$, and $10^{12.8}\Msun$ \citep{Guo2010}. We also tested to see if there was any significant difference in the time that halos spend inside a host as a function of their mass. We find that all the halos with $M\sub{peak}<10^{12}\Msun$ have a similar distribution of time spent in their host (the mean time spent in their host is 2.7\Gyr), while halos with $M\sub{peak}>10^{12}\Msun$ tend to spend shorter times in their hosts ($\sim$2.0\Gyr).

In Figure \ref{fig:f3}, we divide the preprocessed sample by the mass of their host prior to infall into the cluster. Host mass is measured at its peak value. In the upper panel, we show the cumulative percentage of preprocessed halos -- that is similar to Figure \ref{fig:f2}, only with a different choice of subsamples. $\sim30\%$ of the cluster members spent time in a host that is more massive than $10^{13}\Msun$ (i.e. group mass or more) for at least one snapshot. This is equivalent to about two-thirds of the preprocessed halos, so in fact, the massive hosts dominate the contribution of preprocessed halos that enter the cluster. Nevertheless, at $\sim16\%$ the contribution of lower mass hosts ($M<10^{13}\Msun$) is certainly not negligible (equivalent to about one-third of the preprocessed halos). In the lower panel, we show the time distribution of preprocessed halos normalized by the total number of cluster members. The area under each colored line is equivalent to the total fraction of halos from each host mass bins (i.e. the y-intercept of the lines in the upper panel). Halos in lower mass hosts tend to spend slightly shorter times in their host. Also, the contribution of massive hosts is greater if we restrict our preprocessed sample to objects that spend longer times in their host.

In panel (a) of Figure \ref{fig:f4}, we plot the percentage of former group members for each cluster against cluster mass at $z=0$ and find no clear relationship. Instead, we find that the percentage is more clearly related to the recent merger history. To quantify the recent growth rate of the cluster, we measure the time it took a cluster to double (see panel (b)) or triple (see panel (c)) its mass in order to reach its current day mass. Thus, a short time period means a more rapid recent growth in cluster mass. With the exception of one cluster, in all cases, we find a clear trend that clusters with a more rapid recent growth resulted in a higher fraction of former host members.

This can be understood physically because, if a cluster has a recent group merger, a large number of group members may have recently joined the cluster, and this also tends to cause a rapid increase in cluster mass. Meanwhile, in the absence of mergers with massive hosts, clusters grow in mass more steadily and the percentage is continuously diluted by the influx of non-host members. We note that the same dependency on recent growth history remains, even if we conduct the same test at other epochs in the Universe (see Appendix \ref{apx:af1}).

\begin{figure}
\figurenum{5}
\phantomsection
\label{fig:f5}
\includegraphics[width=0.5\textwidth]{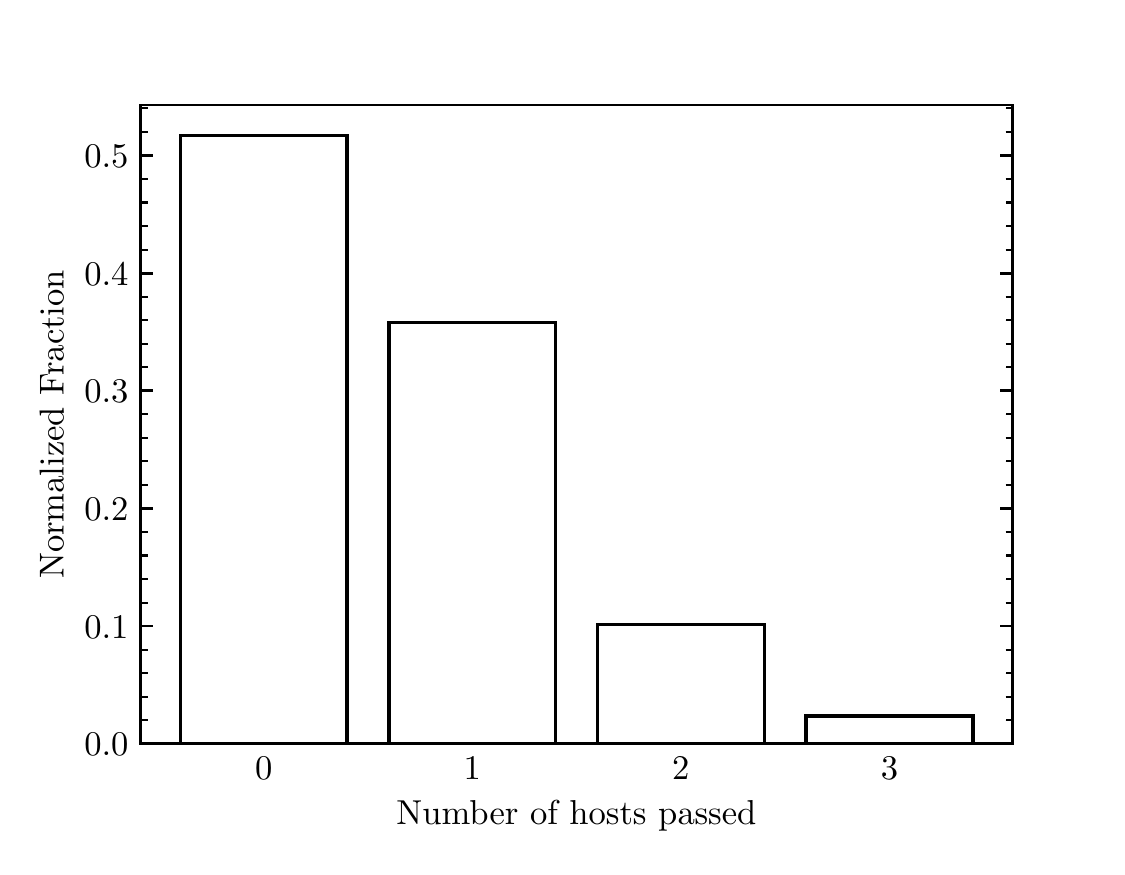}\caption{The number of hosts a halo passes through before becoming a cluster member. Nearly half of the cluster members join the cluster without a host. But of those that were in a host, three quarters had just one host, while one quarter had more than one host prior to cluster infall.}
\end{figure}

Figure \ref{fig:f5} shows the number of distinct hosts that halo met before their cluster infall. As indicated previously, $\sim52\%$ of the cluster member never entered a host before joining the cluster. $\sim12\%$ of the cluster member pass through more than two hosts before infall (i.e. a host that became a satellite of another host before joining the cluster). This means that roughly one-quarter of preprocessed halos actually had multiple hosts prior to cluster infall.

\subsection{Tidal mass loss before cluster infall}

We now define several parameters to describe the amount of tidal mass loss suffered by cluster members. The fractional mass loss from preprocessing is denoted $f\sub{pp}$ and is defined as the fraction of mass lost between peak mass ($m\sub{peak}$) and the mass when the halo became a cluster member ($m\sub{infl}$) (Equation \ref{eq7}). The total fractional mass loss, $f\sub{tot}$ is measured from the peak mass to the mass at $z=0$ ($m_{z=0}$) (Equation \ref{eq8}). We note that some hosts may survive after cluster infall and continue to cause tidal stripping, even after the galaxy crosses the virial radius (post-processing). But, since it is difficult to distinguish between cluster stripping and post-processing stripping, we only consider the mass loss outside the cluster in our pre-processing mass loss fraction. Thus, our measured fractional mass loss from preprocessing may be considered a lower limit on the true mass loss arising from tidal interactions with hosts.

\begin{equation}\label{eq7}
f\sub{pp}=1-\frac{m\sub{infl}}{m\sub{peak}}
\end{equation}

\begin{equation}\label{eq8}
f\sub{tot}=1-\frac{m_{z=0}}{m\sub{peak}}
\end{equation}

To avoid our results being strongly influenced by noisy fluctuations due to errors in measuring the halo mass, we first smooth the mass evolution with a 1-dimensional Gaussian filter with a bandwidth of 2.5 snapshots, which corresponds to $\sim0.2\Gyr$ in time.

\begin{figure}
\figurenum{6}
\phantomsection
\label{fig:f6}
\includegraphics[width=0.5\textwidth]{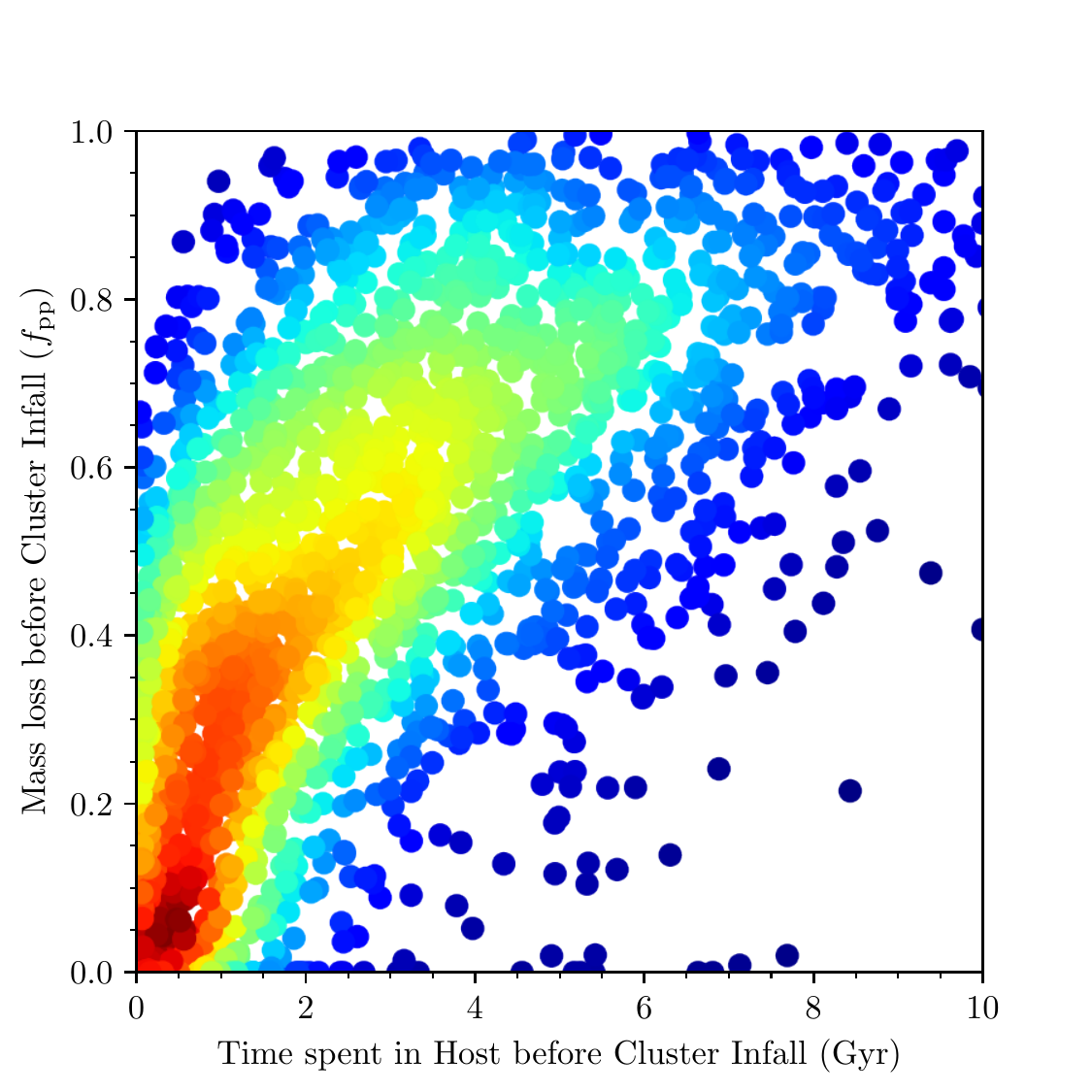}\caption{Fractional mass loss of halos ($f\sub{pp}$) as a function of time spent in their host before cluster infall. Color indicates the number density of points on the figure. Halos that spend longer in their host lose a greater fraction of their mass. But the scatter in the correlation is large, suggesting the presence of secondary parameters that additionally influence the rate of halo mass loss. As the instant of cluster infall differs between galaxies, each data points value is measured at a different epoch of the Universe.}
\end{figure}

Figure \ref{fig:f6} shows the overall relation between the time spent in a host and the fractional mass loss from preprocessing ($f\sub{pp}$) for the sample which has fallen into a cluster. The fractional mass loss that occurred prior to cluster infall is strongly dependent on the time a satellite spent in their host. This indicates that the amount of tidal stripping a satellite suffers is primarily determined by how long they have spent in their host. However, we note that this figure contains halos with a range of properties and includes objects that fell into the cluster very early on, and also other objects that fell into the cluster only recently.

\begin{figure*}
\figurenum{7}
\phantomsection
\label{fig:f7}
\includegraphics[width=1\textwidth]{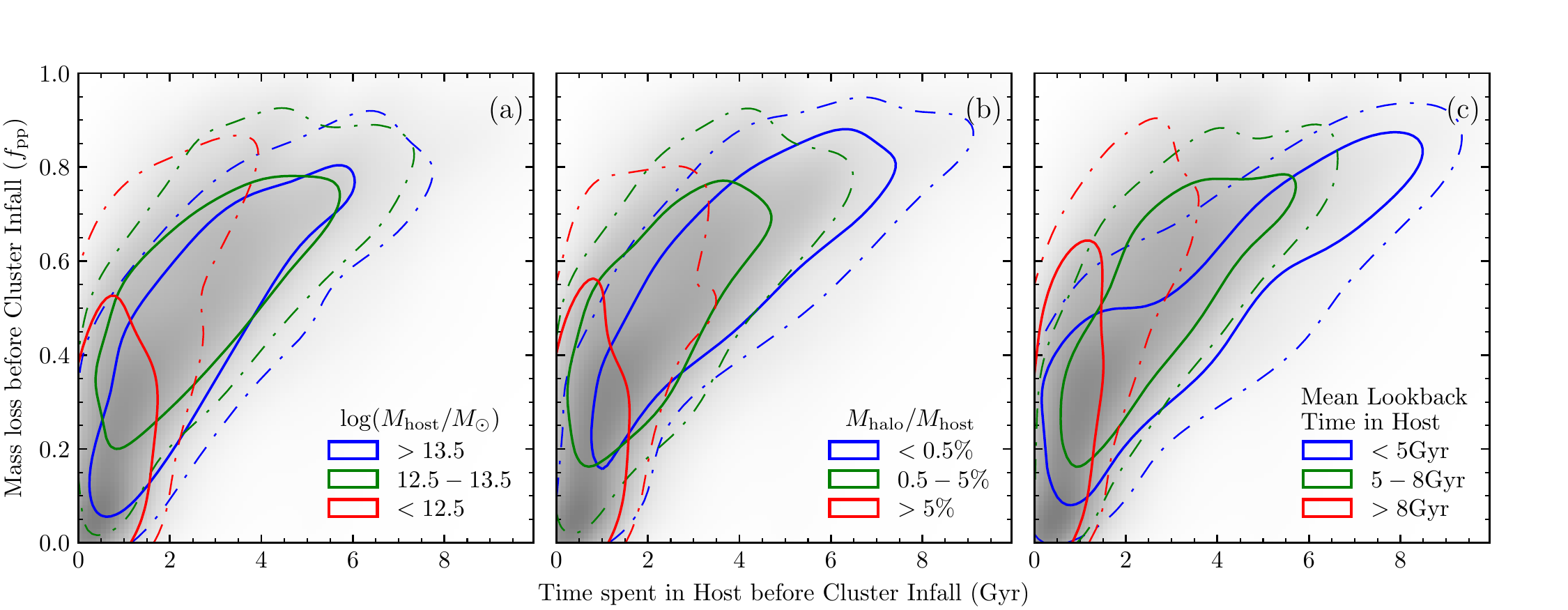}\caption{
Fractional mass loss of the satellites in hosts as a function of time spent in their host. Subsamples are divided by (a) the mean mass of the host, (b) the mass ratio between host and halo, and (c) the mean lookback time that the halo was in its host. The thick solid line and dash-dotted line enclose confidence regions of the halo distribution on the figure ($0.5\sigma$ and $1\sigma$ respectively). The gray smooth contours in the background are the number density of the total sample of host members. In panels (b) and (c), the mass loss rate depends on the mass ratio and the lookback time. In panel (a), the mass loss rate is fairly independent of host mass, except in the least massive hosts (red) where the mass ratio also plays a role. Each data point used to construct this figure is measured at the moment of cluster infall.}
\end{figure*}

\begin{figure*}
\figurenum{8}
\phantomsection
\label{fig:f8}
\includegraphics[width=1.0\textwidth]{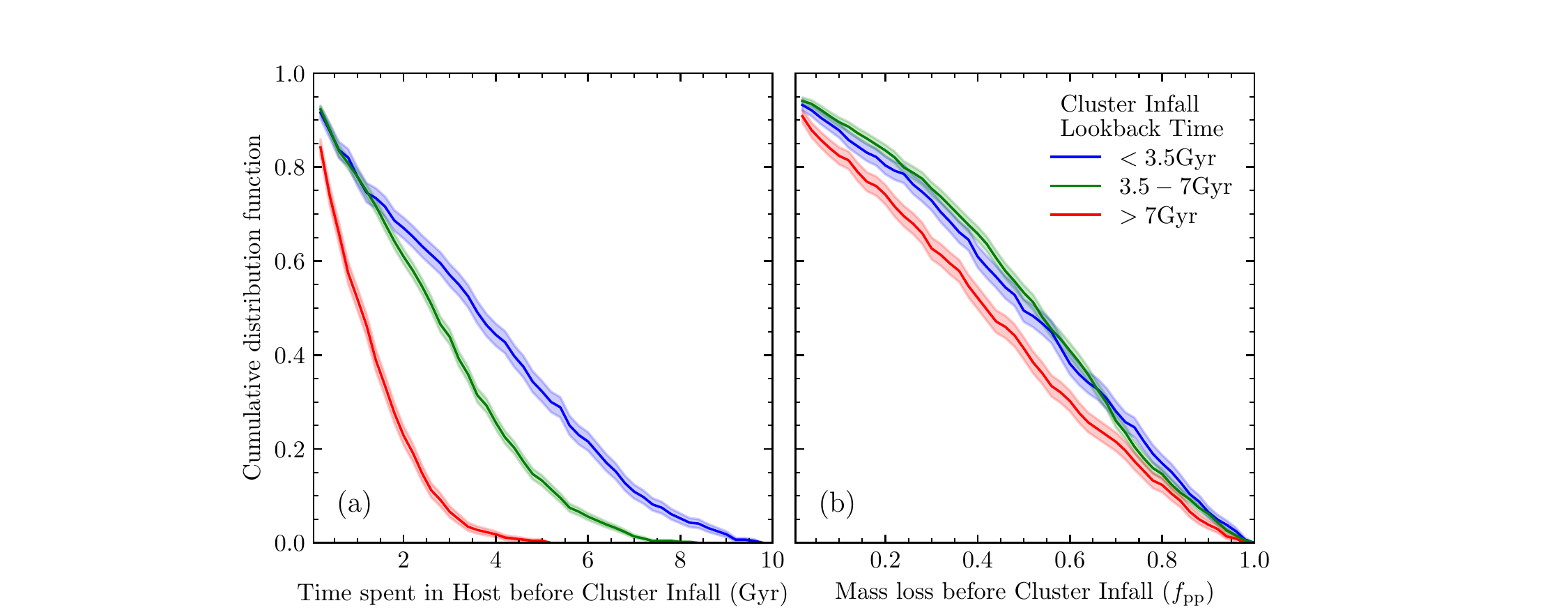}\caption{The normalized cumulative distribution of halos that were previously a satellite of a host as a function of (a) the time spent in hosts before cluster infall, and (b) the fractional mass loss before cluster infall. The samples are sub-divided by cluster infall lookback time. Those halos that fell into the cluster at early epochs have limited time in their hosts. However, the more destructive nature of early hosts (see panel (c) of Figure \ref{fig:f6}), results in a similar amount of mass loss by preprocessing regardless of the time of infall. The shaded regions indicate $1\sigma$ binomial uncertainties.}
\end{figure*}

We will now consider three separate parameters; host mass, the mass ratio between the host and its satellite, and the mean epoch of the Universe that the satellite was a member of its host. The dependency on all three parameters is shown in Figure \ref{fig:f7}. In each panel, one parameter is varied (the parameter is indicated in the legend) by taking a subsample of the total sample. The thick solid line and the dash-dotted line indicates $0.5\sigma$ and $1\sigma$ confidence region of the distribution respectively.

Panel (a) indicates that satellites lose mass at about the same rate in massive, cluster mass hosts as they do in intermediate, group-mass hosts. We see no indication of stronger tidal stripping to occur in more massive hosts. In fact, we see a more rapid mass loss rate only when the host mass is low. However, we suspect that there is actually little dependency on host mass and that our lowest mass host subsample is actually being more influenced by our second parameter, the mass ratio, than by the host mass itself. This is because, as we impose a halo mass cut, our low mass hosts tend to have high mass ratios with their satellites.

Panel (b) of Figure \ref{fig:f7} shows that the gradient of the trend is a clear function of the mass ratio between the host and the satellite. This mass ratio is obtained by averaging the value over the time the satellite is inside the host. The contours of all three subsamples are well-separated. Satellites with a high mass ratio of $M\sub{halo}/M\sub{host}$ tend to lose more mass in a fixed amount of time. This could be consistent with the picture that the orbits of more massive galaxies tend to decay faster within their host due to dynamical friction, resulting in a faster mass loss \citep{Boylan-Kolchin2008}. We also tried measuring the mass ratio when the satellite first becomes a member of the host to found qualitatively very similar results.

In panel (c) of Figure \ref{fig:f7}, the sample is divided by the mean lookback time when the satellite was a member of the host. It is interesting that there is such a clear trend with this parameter. For example, the subsample of halos with a mean lookback time when they were in their host of more than 8\Gyr ago (red contour) seems to have lost more mass in a given time the other subsamples which were in the host in recent times. This implies that, at higher redshift, satellites lost their mass more rapidly in hosts. We also tried dividing the sample by the lookback time when the halo fell into the cluster and splitting into subsamples depending on the epoch of the Universe at which that data point was measured (instead of just using mass loss fraction at the moment of cluster infall). In both cases, we found the results were qualitatively similar to those shown in panel (c). This strongly supports our conclusion that \textit{hosts at higher redshift were significantly more destructive than their low redshift counterparts}. We will discuss this further in Section \ref{sec:prog_bias}.

In Figure \ref{fig:f7}, we only control for one parameter at a time. However, there is a potential risk that between subsamples, the distribution of another influential parameter could be correlated with the first, and they vary at the same time. Therefore, we also tried separating our total sample into 9 panels to check if the effect of the parameters are truly independent of each other, controlling for all three parameters (host mass, mass ratio and mean time in hosts) simultaneously (see Appendix \ref{apx:af2}). By dividing the sample into 27 subsamples, some of our subsamples became quite small in number, and so noisy. However, even when we attempt to control for the host mass, and the mass ratio of the satellite to its host, the effect of the changing epoch remains clearly significant and is one of the strongest factors controlling the mass loss rate of the satellites within their hosts. Overall, we found that the results shown in Figure \ref{fig:f7} are actually a good summary of the behavior seen when controlling for all parameters at the same time. These results demonstrate that a lot of the spread in the correlation seen in Figure \ref{fig:f6} is primarily the result of two parameters varying within the plot - the mass ratio of a satellite with its host and, in particular, the epoch at which the satellites were members of their host. 

Although hosts at earlier epochs of the Universe may be more destructive to their satellites, they also have a limited window of time to act on their satellites before they become members of the cluster. This is shown in panel (a) of Figure \ref{fig:f8} where we plot the normalized cumulative distribution of the time spent in a host before becoming a cluster member. The sample is divided by the lookback time of cluster infall. Those satellites that join the cluster early on (red distribution) tend to spend only a short time period in their host, but are also more likely to be affected by early hosts compared to other subsamples. Panel (b) of Figure \ref{fig:f8} demonstrates that the earlier hosts are so destructive that they can compensate for the limited time they get to affect their satellites. Here we plot the normalized cumulative distribution of fractional mass loss occurring before joining the cluster. The satellites which enter the cluster early lose almost equal fractions of their halo mass, despite the much shorter time they are satellites of their host. In summary, the more destructive nature of groups at earlier epochs means that halos falling into the cluster inside of a host suffer similar amounts of tidal mass loss, regardless of the cosmic age at which they are accreted.

\begin{figure}
\figurenum{9}
\phantomsection
\label{fig:f9}
\includegraphics[width=0.5\textwidth]{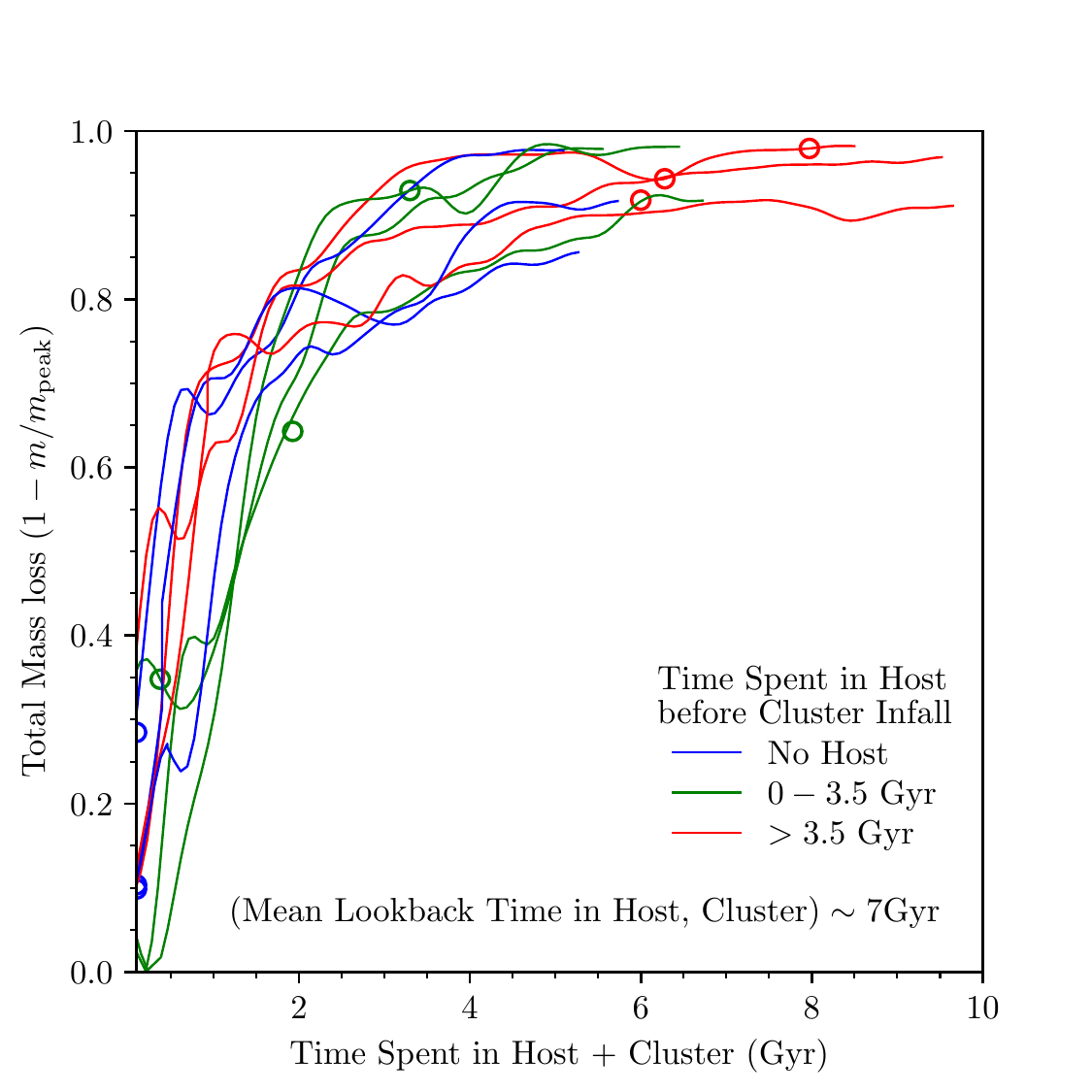}\caption{
Representative examples of the mass loss evolution of 9 halos as a function of the combined time spent in their host and the cluster. The moment of cluster infall is marked with an open circle. Symbol and curve color indicates the time spent in a host before cluster infall (see key). Blue lines indicate halos which fall directly into the cluster. We fix the mean lookback time in host and cluster to be approximately 7~Gyr, to rule out the bias due to redshift seen in panel (c) of figure \ref{fig:f7}. Regardless of when the cluster infall occurs, the mass loss of halos shows a similar overall trend, with an initially steep increase, followed by a slow down as mass loss increases.}
\end{figure}

Since all kind of hosts, including the cluster, groups, and lower mass hosts can cause a mass loss in their satellites, we now consider the combined mass loss that took place both before and after the cluster infall. In Figure \ref{fig:f9}, we look at some representative mass loss curves of some individual halos. The y-axis is the total fractional mass loss measured with respect to the peak mass. The total time spent in both hosts and the cluster combined is shown on the x-axis. We indicate the moment of cluster infall of each halo with an open circle, and color the curve and symbols by the time they spent in their host before cluster infall (the x-axis position of the open circle). To rule out the influence of the epoch of the Universe (that we saw in panel (c) of Figure \ref{fig:f7}), we fix the mean lookback time that the halo was in its host (or in the cluster) to be approximately 7\Gyr. \citet{Joshi2017} discovered that preprocessed cluster members suffer slower mass loss inside the cluster than halos that were isolated prior to cluster infall. We confirm their result in Figure \ref{fig:f9}, with the preprocessed members (green and red) suffering significantly slower mass loss after joining the cluster (see the curve gradient to the right of the symbol) compared to members that directly fell into the cluster as single halo (blue). However, an additional point we can see in Figure \ref{fig:f9} is that, overall, the mass loss curve has a very similar shape, independent of the time spent in their host. This means that the change in the gradient of the mass loss curve is not due to the fact that the halos are in the cluster. Rather, it is a natural result of the fact that the mass loss rate from tidal stripping depends sensitively on the remaining mass of the satellite. Physically this might be understood by the fact that, thanks to the concentrated density profile of a dark matter halo, the outermost low-density halo is quickly and easily stripped, meanwhile the dense innermost regions of the halo are much more robust to tidal stripping.

\section{Discussion}\label{sec:discussion}
\subsection{The significance of Preprocessing}
On average, we found nearly $50\%$ of the cluster member halos were previously within another host before joining the cluster. The fraction has no trend with cluster mass. Furthermore, the fraction is also independent of the stellar mass of the satellites below $M_*\sim10^{10.5}\Msun$\footnote{Although we only consider dark matter halo evolution in this study, we have also measured the stellar mass of the galaxies.}. However, these results seem contradictory from earlier study \citep{McGee2009} which found dependencies on both cluster and galaxy mass. This discrepancy likely arises because we do not limit ourselves to considering preprocessing in hosts with group masses only and consider the contribution from lower mass hosts down to mass limit as well. Thus, as a result of self-similarity, there is much less dependence on cluster or satellite mass in our results. Second, our criterion for the definition of satellite membership is somewhat different from those in other studies (see Equation \ref{eq1}--\ref{eq6} and Figure \ref{fig:f1}). Also, some of the previous studies included the central halo of the group as a preprocessed member whereas we exclude them. In order to better compare with previous studies, we also tried to match their criterion. We limited our study only to group-mass hosts ($>10^{13}\Msun$) and considered satellites that are within one virial radius instead of using our criterion, while also included group centrals in the preprocessed population. From this, we obtained $\sim28\%$ as a fraction of preprocessed galaxies $M_*\sim10^{9}\Msun$ in clusters of $\sim10^{14}\Msun$, which is fully consistent with the results of \citet{McGee2009} and \citet{DeLucia2012}.

However, we note that it is better to consider hosts of all masses when trying to understand the contribution of preprocessing to the cluster population. According to the top panel of Figure \ref{fig:f3}, the relative contribution between hosts with mass $<10^{13}\Msun$ and mass $>10^{13}\Msun$ is about 1:2. Therefore, the low mass host contribution is certainly not negligible by number. Moreover, we found that the strength of tidal stripping is actually quite independent of mass. It may even be slightly stronger in the lower mass hosts, although it seems this may actually be more the actions of artificially enhanced mass ratios between the satellite and its host due to the minimum halo mass limit that we impose, rather than because of the host mass itself.

It is often stated that dwarf galaxies can be used as sensitive probes of their environment due to their shallow potential wells \citep{Haines2007, Pustilnik2011, Leaman2013, Lisker2013, Boselli2014a, Weisz2014}. While this is likely the case for some environmental mechanisms such as ram pressure stripping \citep{Mori2000, Mayer2006, Sawala2012, Bahe2015}, our results suggest this is not the case for tidal stripping. Given we find faster mass loss for higher $M\sub{sat}/M\sub{host}$ ratio, then in a fixed host mass, it is actually \textit{more massive satellite galaxies that suffer faster mass loss}. Conversely, this dependence on mass ratio means that, for a fixed mass of the galaxy, \textit{a group host is in fact more destructive than a cluster}.

\begin{table}[!ht]
\begin{threeparttable}
\caption{Mean fractional mass loss of the surviving cluster members} % title name of the table
\label{table:mean_loss}
\centering % centering table
\renewcommand{\arraystretch}{1.25}
\begin{tabularx}{0.45\textwidth}{@{}l >{\centering\setlength\hsize{1.2\hsize}\arraybackslash}X *2{>{\centering\setlength\hsize{0.4\hsize}\arraybackslash}X}@{}}
\toprule % inserting double-line
 Sample & Number & $\langle f\sub{pp} \rangle$ & $\langle f\sub{tot} \rangle$ \\
 & & (\%) & (\%) \\
\midrule % inserts single-line
No Host\tnote{a} & 1231 & 11 & 47 \\
In Host\tnote{b} & 1156 & 44 & 67 \\
Total & 2387 & 28 & 57 \\
\bottomrule % inserts single-line

\end{tabularx}
\begin{tablenotes}
\item[a] Satellites that were not previously in hosts
\item[b] Satellites that were in hosts before cluster infall
\end{tablenotes}
\end{threeparttable}
\end{table}

We now try to quantify the significance of preprocessing for the total mass loss suffered by cluster members. (See Table \ref{table:mean_loss}) The total mean fractional mass loss of the surviving cluster members from mass peak to $z=0$ was $\langle f\sub{tot}\rangle \sim57\%$. We found these halos already lost on average  $\langle f\sub{pp}\rangle \sim28\%$ of their mass before reaching the cluster, which suggests that the contribution of preprocessing is roughly half (28/57) of the total mass loss of the surviving cluster halos. However, if we consider cluster members that were formerly satellites of hosts, the mean fractional mass loss increases to $\langle f\sub{tot}\rangle \sim67\%$, and $\langle f\sub{pp}\rangle \sim44\%$. Thus, preprocessing results in almost two-thirds of the total mass loss for cluster members that were previously in a host. Despite differences in the definition of satellites and the mass cut used, the fraction is in good agreement with that quoted in \citet{Joshi2017}. We also calculate these values for halos that were without a host until cluster infall (labeled `isolated' in the Table). They recorded an average mass loss of $\sim11\%$ before entering the cluster. The fact that the value is non-zero suggests that a small amount of mass loss can occur even before becoming a cluster member, and in the absence of a host perhaps as a result of fly-bys or the background potential gradient from the large-scale structure, although it is clearly minor in comparison to mass loss occurring inside hosts.

As a galaxy's stellar component is much more compact than its dark matter halo, and tidal stripping tends to occur preferentially in the outskirts, large amounts of halo mass loss can be stripped before any stars are removed. Using the same set of hydrodynamic zoom-in cluster simulations as used in this study, \citet{Smith2016} states that typically the loss of $\sim 80\%$ of the halo mass results in only $\sim 10\%$ stellar mass loss. Therefore, we decided to choose a `high mass loss' sample of cluster members, that have lost more than $80\%$ of the halo mass at $z=0$, and investigate the role of preprocessing for them. In total, $\sim 20\%$ of the cluster members are included in the high mass loss sample. On average, preprocessing alone is not strong enough to result in stellar stripping -- only $\sim 5\%$ of the cluster halos lost this amount of mass prior to cluster infall. This is because, even though nearly half of the cluster members have spent time in a host before cluster infall, most of these halos spent only a short time in their host, resulting in only a small fraction of halos with strong tidal stripping. However, preprocessing in combination with cluster tides can result in a significant tidal mass loss -- we found $\sim 74\%$ of the high mass loss sample was previously in a host before cluster infall, and so galaxies that suffered preprocessing strongly dominate the high mass loss sample. Furthermore, they lost on average $\sim 62\%$ of their mass before becoming a cluster member. Overall, these results imply that preprocessing in a host can conveniently extend the timescales that tides can act on a halo over its lifetime. By falling first into a host, they are given an opportunity to extend the time they spend suffering tidal stripping beyond that they would experience if they had fallen directly into the cluster. While it is possible for non-preprocessed objects to reach similar levels of mass loss (i.e. by spending sufficient time in the cluster) such cases are rare in the hierarchical clustering paradigm, as an insufficient number of objects fall into the cluster so early. A strong mass loss is much easier and more common with the aid of preprocessing. When combined with further tidal stripping in the cluster, the accumulated damage can be sufficient to result in a significant stellar stripping.

\subsection{`Progenitor bias' of group-mass hosts}\label{sec:prog_bias}
On panel (c) of Figure \ref{fig:f7}, we present the important result that, at earlier epochs, hosts were more destructive to their satellites, which is also the case for group-mass hosts. This can be thought of as a `progenitor bias' because groups in the early Universe and today's groups could have evolved from distinctive progenitors, thanks to the continuous process of hierarchical growth. This may make estimating the contribution of preprocessing to the current day cluster population based on today's groups a risky endeavor. Indeed, as earlier groups are more destructive, we may underestimate the significance of preprocessing in this way. This increases the relative importance of the tidal stripping occurring in early groups, in a manner which is consistent with the results of \citet{Lisker2013}.

\begin{figure}
\figurenum{10}
\phantomsection
\label{fig:f10}
\includegraphics[width=0.5\textwidth]{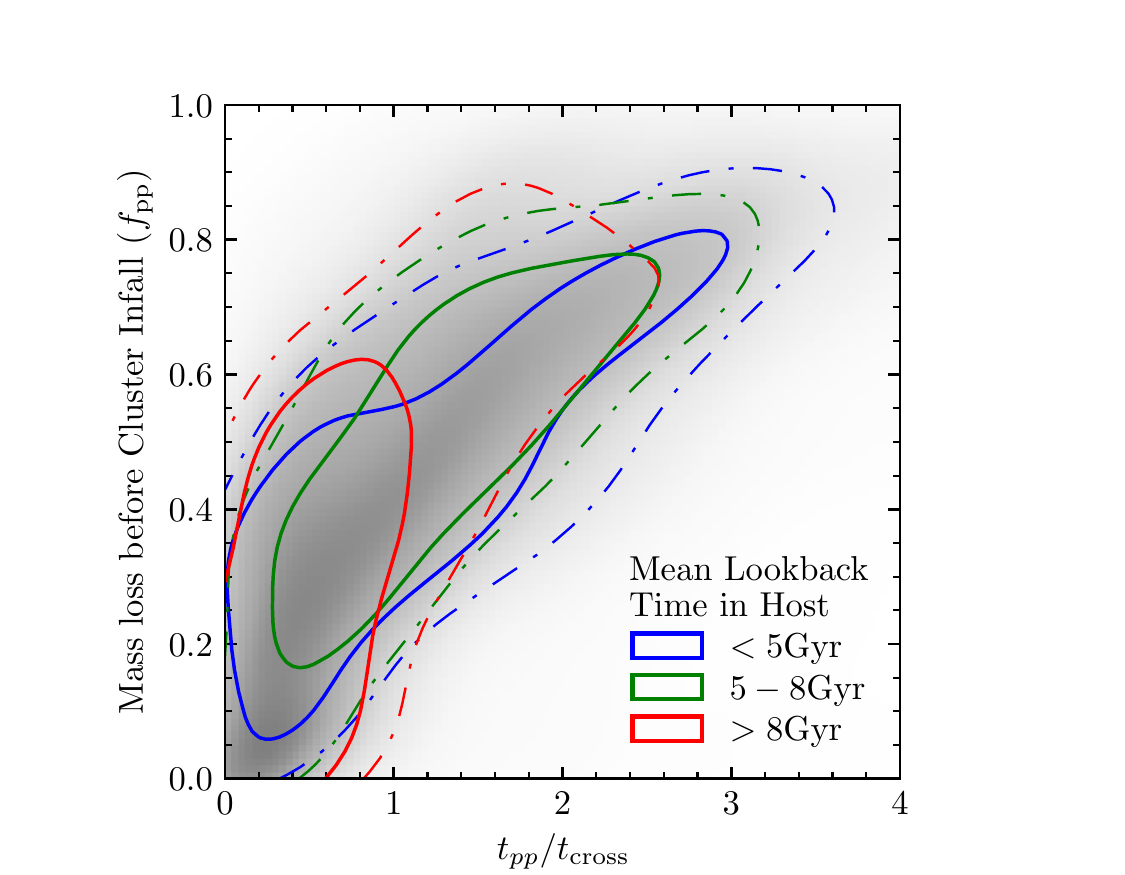}\caption{
The fractional mass loss of halos that occurs before cluster infall. The x-axis is the number of crossing time that the halo spent in its host. This figure can be compared with panel (c) of Figure \ref{fig:f7} (where the time axis is not normalized). By normalizing by crossing time, the slope of 3 contours become more similar. This implies that a large amount of the progenitor bias effect we have seen previously is due to variations in the crossing time with redshift.}
\end{figure}

It is not entirely clear what physical properties of the early hosts make them more hostile to their satellites. However, we can expect certain possibilities. The group halos are likely to have higher mean densities at earlier times, and more intense tidal fields. The higher mean densities of early groups will also naturally result in shorter crossing timescales compared to today's groups. Assuming that the crossing time, $t\sub{cross}$, is twice that of the free-fall time we can write, 

\begin{equation}\label{eq9}
t\sub{cross}=2t\sub{ff}=\sqrt{\frac{3\pi}{8G\rho}}
\end{equation}

\noindent
where $\rho$ is the density within the virial radius of the halo, which is approximately $\sim200$ times the critical density of the Universe at that epoch.

Thus, as the critical density of the Universe is higher at earlier epochs, then $t\sub{cross}$ is shorter, which could result in a faster mass loss through more frequent pericenter passages with their hosts. To test this, we measure the fractional mass loss as a function of the number of crossing times (calculated using Equation \ref{eq9}) that each halo spent inside their host before cluster infall. The result is shown in Figure \ref{fig:f10} and can be directly compared with panel (c) of Figure \ref{fig:f7}. It can be seen that the differences between the 3 subsamples are greatly reduced, which indicates that the crossing time is a key factor driving the progenitor bias.

However, there is still a systematic shift towards the faster mass loss in earlier hosts requiring the role of some additional effects. We investigated if the evolution of halo central densities with time could be responsible, by measuring the mean density of halos within their scale radius. However, although we see evolution of the central density with time, we found that the ratio of host-to-satellite density evolves very little. Thus even if hosts are denser at a particular epoch resulting in stronger tides, their satellites are similarly denser and thus more robust to those tides. However, other alternative processes may be responsible. In the early Universe, groups may evolve through a period of more violent and frequent mergers, which could enhance tidal stripping and the chance of close encounters between satellites. And the satellites themselves may be less relaxed than today's counterparts, thus potentially more vulnerable to tidal stripping. Also, \citet{Wetzel2011} finds that orbit of satellites on first infall into a host is a function of redshift, with more radial and plunging orbits occurring at higher redshifts, and eccentricity of the initial orbit is known to be a critical factor determining the mass loss rate \citep{Boylan-Kolchin2007, Villalobos2012}.

\section{Conclusion}\label{sec:conclusion}
Using a set of cosmological hydrodynamic zoom-in simulations of clusters, we attempt to better understand the significance of preprocessing for the cluster population. We focus on the fraction of the cluster population that were preprocessed and study the amount of tidal stripping that their dark matter halos suffer. We also determine the key parameters controlling the amount of mass loss satellites suffer inside their hosts. Our zoom-in simulations are well suited to tackling this problem as, by zooming on clusters, we achieve comparable resolution with modern-day cosmological hydrodynamic simulations \citep{Dubois2014, Vogelsberger2014, Schaye2015}, but with improved number statistics in dense regions. 

Our study has several unique features compared to previous studies on this topic. We apply a new criterion to classify which galaxies are members of hosts (see Equation \ref{eq1}). This criterion effectively captures backsplash galaxies which are bound to their host but currently found beyond the virial radius of the group. It also effectively excludes unbound, high-velocity fly-bys. In addition, we also do not limit ourselves to only consider group-mass hosts. Instead, we include hosts of all masses within our sample. This enables us to uncover the significant contribution of lower mass hosts to the tidal mass loss resulting from preprocessing. Our results can be summarized as follows.

\renewcommand{\labelenumi}{(\roman{enumi})}
\begin{enumerate} 

\item We find a large fraction ($\sim48\%$) of cluster members have spent some time as a satellite of a host prior to becoming a cluster member. However, the fraction falls to only $\sim4\%$ if we require them to have spent at least 6\Gyr in their host. Thus, faster acting environmental mechanisms in a host may affect a larger fraction of the cluster population.

\item There is a large scatter in the fraction of cluster members that were formerly in a host from cluster-to-cluster with no clear dependency on cluster mass. Instead, we find the fraction depends most clearly on the recent mass growth of the cluster, with rapid recent growth enhancing the fraction.

\item With regards to mass loss, hosts less massive than $10^{13}\Msun$ should not be neglected. Close to one-third of all the preprocessed cluster members came from such a low mass host (18$\%$ of the total cluster members).

\item Tidal stripping due to preprocessing is very significant for the surviving cluster population. Indeed, on average half of the total dark matter mass loss of today's cluster population occurs prior to cluster infall. For cluster members that have suffered a large amount of tidal stripping\footnote{Defined as having lost $>80\%$ of their peak mass by $z=0$, which is sufficient to cause significant stellar stripping \citep{Smith2016}.}, $\sim74\%$ were members of a host before joining the cluster. By infalling into a host before reaching the cluster, halos can extend the time they suffer tidal mass loss beyond that they would typically experience if they fell directly into a cluster (e.g. see Table \ref{table:mean_loss}).

\item We find that, in all cases, halo mass loss is a clear function of time spent in a host. However, other parameters can enhance the rate of mass loss. We do not find a clear dependency on the mass of the host. But when a satellite and its host have more similar mass, there is faster mass stripping. This means that for a fixed mass of host, it is \textit{massive satellites that suffer the fastest mass loss}. Conversely, at fixed satellite mass, \textit{group mass hosts are more destructive than cluster hosts}.

\item We clearly see that \textit{hosts at higher redshift are more destructive to their satellites than their low redshift counterparts} (see panel (c) of Figure \ref{fig:f7}). A large part of this `progenitor bias' is driven by variations in crossing time with the epoch of the Universe. The effect is sufficiently strong that preprocessed halos suffer nearly equal quantities of mass loss prior to cluster infall, independent of their time of infall.

\end{enumerate}

Certain classes of object, such as ultra-compact dwarfs, and perhaps some globular clusters, may have their origin in the heavy tidal stripping of more massive progenitor systems \citep{Lee1999, Bekki2001, Bekki2003, Pfeffer2014}. Pre-processing in hosts, prior to cluster infall, could play an important role in producing heavily tidally stripped objects in clusters at redshift zero. We have shown that heavy mass loss requires long periods of time inside the potential well of a more massive system. By falling into a host at early times, galaxies that will end up in the cluster can dramatically extend the time they suffer tidal stripping and may also suffer more rapid tidal stripping in such an environment.

In this study, we have primarily focused on the mass evolution of dark matter halos, which is the most extended, and hence sensitive component, of a galaxy to tidal stripping. However, this makes it difficult to compare our results directly with observations. Because tidal stripping preferentially occurs in the outer part of the halo, only strong tidal stripping will have an impact on the observable components. In the future, we plan to better explore the observable implications of preprocessing for the cluster population, considering galaxy stellar masses instead of halo masses, and studying the impact of hosts on the star formation rates of their satellites. We will also consider how heavily preprocessed galaxies distribute themselves spatially in clusters today and their association with substructure. These results will be presented in an upcoming paper.

\acknowledgments
We thank the anonymous referee for providing many useful comments that have helped to improve the paper. S.K.Y. acknowledges support from the Korean National Research Foundation (NRF-2017R1A2A1A05001116). This study was performed under the umbrella of the joint collaboration between Yonsei University Observatory and the Korean Astronomy and Space Science Institute. The supercomputing time for the numerical simulations was kindly provided by KISTI (KSC-2014-G2-003). We acknowledge support by the Australian Government through the Australia-Korea Foundation of the Department of Foreign Affairs and Trade ("Galaxy evolution across environments: linking Australian observatories and Korean simulations" - AKF00601). Parts of this research were conducted by the Australian Research Council Centre of Excellence for All Sky Astrophysics in 3 Dimensions (ASTRO 3D), through project number CE170100013. As the head of the group, S.K.Y. acted as the corresponding author.

\bibliographystyle{mnras}
\bibliography{references}

\appendix
\section{A. Evolution of preprocessed member fraction vs mass doubling time relation.}\label{apx:af1}
\includegraphics[width=1.0\textwidth]{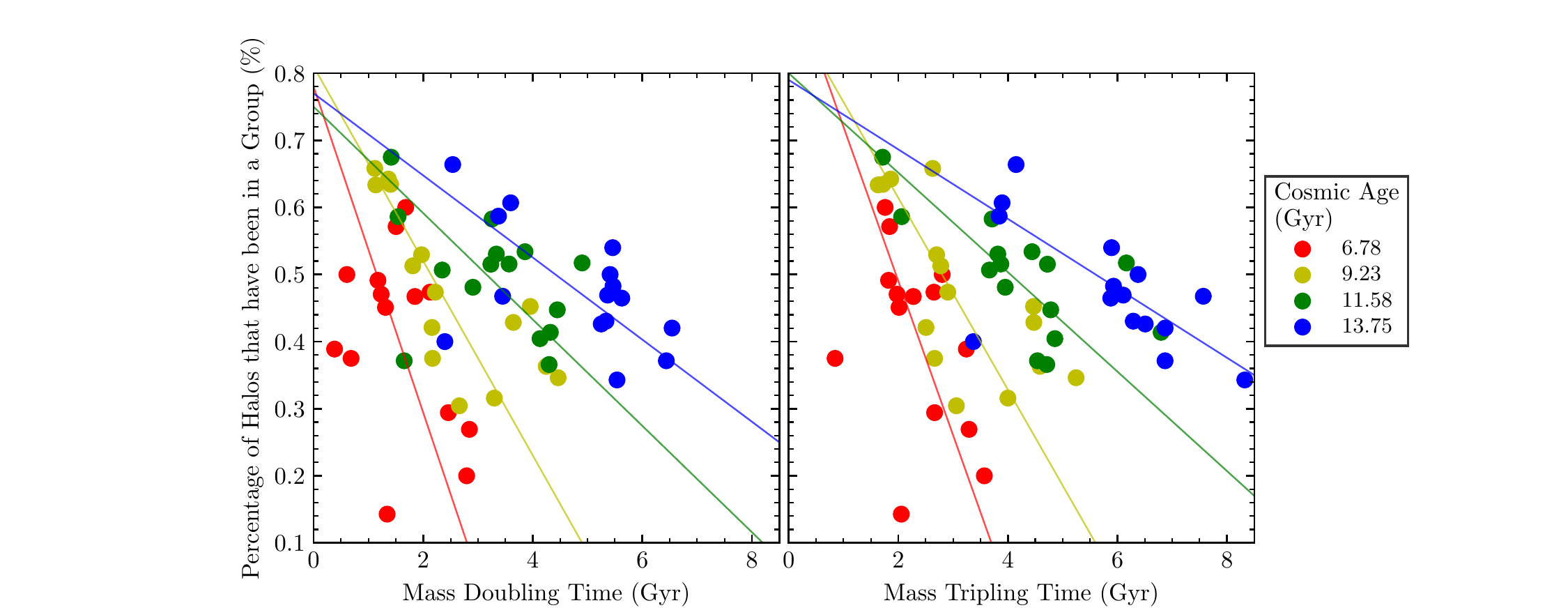}
In this figure, we extend the results shown in Fig \ref{fig:f4} in panels (b) and (c), where we only considered the $z=0$ results. Now we consider other epochs of the Universe, as indicated by symbol color (see key). Lines are linear fits to the data points. We find that the relationship between the percentage of preprocessed members and the recent mass growth history is the same as in Figure \ref{fig:f4} at any epoch, and thus there is nothing unique about the cluster state at $z=0$. However, at earlier epochs in the Universe, the dependency on recent mass growth history becomes even stronger as the available time to increase the cluster mass by a given factor becomes more and more limited by definition.

\section{B. Mass loss vs time relation of group satellites: Full parameter study}\label{apx:af2}
\includegraphics[width=0.98\textwidth]{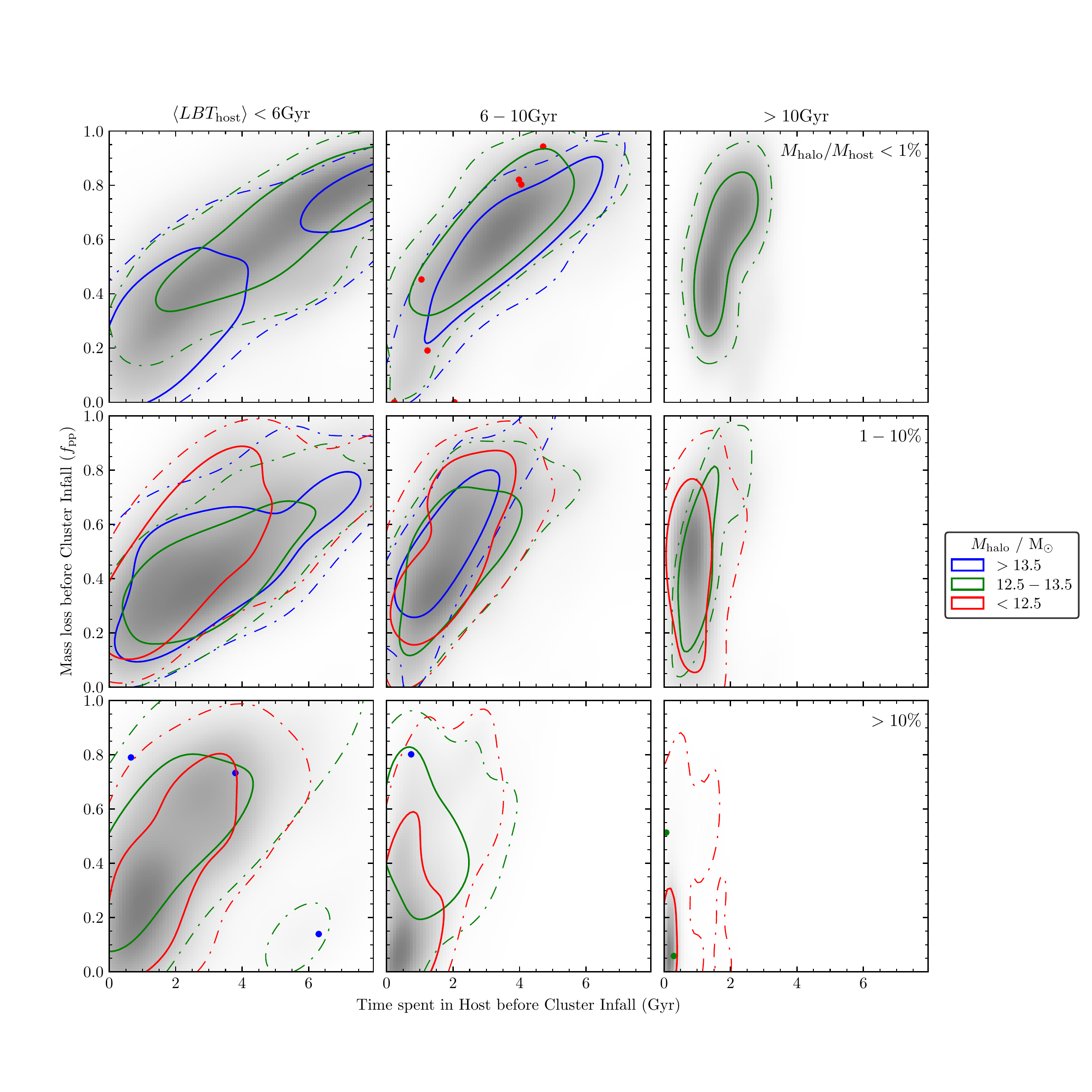}
In this figure, we attempt to better separate the impact of the various parameters in Fig. \ref{fig:f7}, by controlling for each parameter in turn. Each panel shows the fractional halo mass loss as a function of time spent in their hosts prior to cluster infall. Halos are divided by host mass and shown as different color contours in each panel (see key). Now we further divide the samples by mean lookback time when the satellite was in the host (compare along the column) and the mass ratio (compare along the row). This further division of the sample can result in quite low number statistics in some subsamples. If the sample becomes less than 15, we present scatter points instead of contours. Typically subsample sizes are in the range 50--300.

Overall, this exercise confirms the main trends seen in Fig. \ref{fig:f7}. As before, increasing the mass ratio and the lookback time results in increasing mass loss rates. However, now we can see that when we better control for the mass ratio, the role of the host mass is weakened. Indeed, these results also demonstrate that the epoch of the Universe is perhaps the strongest controlling parameter controlling the mass loss rate, when we control for the mass ratio.

\end{document}